\documentclass[prd,aps,preprint,amsmath,amssymb,eshowkeys,nofootinbib]{revtex4-1}
\pdfoutput=1
\usepackage{verbatim,graphics,graphicx,color,slashed,textcomp,bbm}
\usepackage[normalem]{ulem} % this one is for cross out text
\usepackage[colorlinks=true,linkcolor=red,urlcolor=blue,citecolor=blue]{hyperref}

\newcommand{\GeV}{\mathrm{GeV}}

\begin{document}

\title{Conformal ${\alpha}$-attractor Inflation with Weyl Gauge Field}
\author{Yong Tang$^{a,b,c}$ and Yue-Liang Wu$^{c,d,e,f}$}
\affiliation{\begin{footnotesize}
	${}^a$School of Astronomy and Space Sciences, University of Chinese Academy of Sciences (UCAS), Beijing, China\\
	${}^b$National Astronomical Observatories, Chinese Academy of Sciences, Beijing, China \\
	${}^c$School of Fundamental Physics and Mathematical Sciences, \\
	Hangzhou Institute for Advanced Study, UCAS, Hangzhou 310024, China \\
	${}^d$International Centre for Theoretical Physics Asia-Pacific, Beijing/Hangzhou, China \\
	${}^e$Institute of Theoretical Physics, Chinese Academy of Sciences, Beijing 100190, China \\
	${}^f$School of Physical Sciences, University of Chinese Academy of Sciences, Beijing, China   \end{footnotesize}}

\begin{abstract}
Conformal scaling invariance should play an important role for understanding the origin and evolution of universe. During inflation period, it appears to be an approximate symmetry, but how it is broken remains uncertain. The appealing $\alpha$-attractor inflation implements the spontaneous breaking of conformal symmetry and a mysterious SO$(1,1)$ global symmetry. To better understand the SO$(1,1)$ symmetry, here we present a systematic treatment of the inflation models with local conformal symmetry in a more general formalism. We find SO$(2)$ is the other possible symmetry in the presence of Weyl gauge field. We also obtain all the analytic solutions that relate the inflation fields between Jordan frame and Einstein frame. We illustrate a class of inflation models with the approximate SO$(2)$ symmetry and trigonometric potential, and find that it can fit the current observations and will be probed by future CMB experiments.
\end{abstract}	
%\date{}

\maketitle

\section{Introduction}
Inflation~\cite{Guth:1980zm,Linde:1981mu,Albrecht:1982wi,Starobinsky:1980te} in the early universe was proposed to provide an attractive solution for some cosmological puzzles, including flatness problem and horizon problem. During the exponential expansion of inflation, the universe was nearly conformal\footnote{In this paper, we use conformal and Weyl symmetry interchangeably since no ambiguity arises here.} invariant, and the breaking of conformal invariance can provide the primordial perturbations that account for the currently observed inhomogeneity and anisotropy~\cite{Mukhanov:1981xt}. However, the exact mechanism of the breaking is still unknown.

The local Weyl or conformal symmetry was originally motivated to unify Einstein's General Relativity (GR) and Maxwell's electromagnetic (EM) theory~\cite{Weyl:1919fi}, although later it turned out that $U(1)$ gauge symmetry correctly describes the EM interaction. Nowadays, the symmetry still stimulates many theoretical and phenomenological studies. And various applications of global or local conformal symmetry have been explored in, for example, induced gravity~\cite{Zee:1978wi, Adler:1982ri, Fujii:1982ms}, gravitational quantum field theory of fundamental interactions~\cite{Wu:2015wwa,Wu:2017urh}, particle physics~\cite{Hur:2011sv, Holthausen:2013ota, Foot:2007iy, Nishino:2009in}, inflation and late cosmology~\cite{Wetterich:1987fm, Cheng:1988zx, Kaiser:1994vs, Wu:2004rs, Ferrara:2010in, GarciaBellido:2011de, Farzinnia:2013pga, Giudice:2010ka, Kurkov:2013gma, Bars:2013yba, Csaki:2014bua, Guo:2015lxa, Kannike:2015apa, Kannike:2015kda, Salvio:2017xul, Ferreira:2018qss, Tang:2018mhn, Barnaveli:2018dxo, Ghilencea:2018thl, Kubo:2018kho, Tang:2019uex, Ghilencea:2019rqj, Ferreira:2019zzx, Gunji:2019wtk, Ishiwata:2018dxg}. 

The appealing $\alpha$-attractor in Refs.~\cite{Kallosh:2013hoa, Kallosh:2013yoa} was proposed as a class of inflation models with spontaneously broken conformal invariance. These models have an attractor point that predicts the same cosmological observables as in Starobinsky's model~\cite{Starobinsky:1980te}. Interestingly, there is an approximate SO$(1,1)$ global symmetry in such models. However, the origin of the SO$(1,1)$ seems mysterious and it is unclear whether there are other possible symmetries for viable inflation models with conformal symmetry. 

In this work, we present a systematic investigation on the inflation models with local conformal symmetry. To be as general as possible, we include the Weyl gauge field in the starting Lagrangian. Our formalism goes back to $\alpha$-attractor when the relevant parameters are specified. We find an approximate SO$(2)$ global symmetry is also possible for viable inflation and present all the analytic, compact solutions that connect the inflation fields between Jordan frame and Einstein frame. A class of inflation models is illustrated with the approximate SO$(2)$ symmetry, which is found to be consistent with current observations. Such models can be probed by the next-generation experiments in cosmic microwave background (CMB). 

This work is organized as follows. In Sec.~\ref{sec:general} we shall first give an overview of the $\alpha$-attractor and establish the conventions and general formalism with Weyl gauge field for our later discussions. Then in Sec.~\ref{sec:analytic} we work out the analytic solutions that connect the inflation fields in the Jordan frame and Einstein frame. We show the solutions can be classified into two categories, trigonometric functions and hyperbolic ones. Afterwards, we provide a viable and testable inflation model in Sec.~\ref{sec:pheno}. Finally, we give our conclusion.

\section{General Formalism}\label{sec:general}
Throughout the paper, we use the metric with a sign convention $(1,-1,-1,-1)$, and the natural unit, $M_{p}\equiv 1/\sqrt{ 8\pi G}=1$. We shall first review the formalism in $\alpha$-attractor and then present the general formalism with Weyl gauge field.

\subsection{$\alpha$-attractor}
To set the stage for our discussions, we first present the essential formalism of the $\alpha$-attractor~\cite{Kallosh:2013hoa, Kallosh:2013yoa}. The starting action of the $\alpha$-attractor is the following one in Jordan frame,
\begin{equation}\label{eq:alpha}
S=\int d^{4}x\sqrt{-g}\left[\frac{1}{2}\left(\phi^{2}R-6\partial_{\mu}\phi\partial^{\mu}\phi\right)-\frac{1}{2}\left(\chi^{2}R-6\partial_{\mu}\chi\partial^{\mu}\chi\right)-V\left(\phi,\chi \right)\right],
\end{equation}
where $R$ is the Ricci scalar, and the two real scalar fields, $\phi$ and $\chi$, are conformally coupled with gravity. Note that the signs in front of the final kinetic terms of $\phi$ and $\chi$ are {\textit{opposite}}, namely $\phi$ has the wrong sign while $\chi$ has the right one. The above action respects the following local conformal/Weyl symmetry,
\begin{align}\label{eq:weyl1}
g_{\mu\nu}\left(x\right) & \rightarrow\bar{g}_{\mu\nu}\left(x\right)=\lambda^{2}\left(x\right)g_{\mu\nu}\left(x\right),\nonumber\\
\phi\left(x\right) & \rightarrow\bar{\phi}\left(x\right)=\lambda^{-1}\left(x\right)\phi\left(x\right),\\
\chi\left(x\right) & \rightarrow\overline{\chi}\left(x\right)=\lambda^{-1}\left(x\right)\chi\left(x\right),\nonumber
\end{align}
where $\lambda\left(x\right)$ is a non-zero function. In $\alpha$-attractor papers~\cite{Kallosh:2013hoa, Kallosh:2013yoa}, the following specific potential was considered, 
\begin{equation}
V\left(\phi,\chi \right)=F\left(\chi/\phi\right)\left(\phi^2 - \chi^2\right)^2,
\end{equation}
so that there is an approximate SO$(1,1)$ global symmetry for $\phi$ and $\chi$, except the breaking term $F\left({\chi/\phi}\right)$ which is an arbitrary function that depends on $\chi/\phi$ only.

Once fixing the condition that breaks the conformal symmetry spontaneously, $\phi^2=1$, we can define a new metric tensor $\bar{g}_{\mu\nu}$ through conformal transformation,
\begin{equation}
\bar{g}_{\mu\nu}\left(x\right)=\Omega^{2}\left(x\right)g_{\mu\nu}\left(x\right),\;
\Omega^2 = 1-\chi^2,
\end{equation}
and use the following identity for Ricci scalar,
\begin{equation}\label{eq:confR}
R=\Omega^{2}\left[\bar{R}+6\bar{g}^{\mu\nu}\partial_{\mu}\ln\Omega\partial_{\nu}\ln\Omega\right]. 
\end{equation}
Then we can obtain the following action,
\begin{equation}
S=\int d^{4}x\sqrt{-\bar{g}}\left[\frac{1}{2}\bar{R}+\frac{3}{\left(1-\chi^2\right)^2}\partial_{\mu}\chi \partial^{\mu}\chi-F\left( \chi \right)\right].
\end{equation}
To normalize the kinetic term, we can define a new field viable $\theta$ by the differential equation,
\begin{equation}
\frac{d\theta}{d\chi}=\frac{\sqrt{6}}{1-\chi^2}\Rightarrow \chi = \tanh \frac{\theta}{\sqrt{6}},
\end{equation}
and rewrite the final action in Einstein frame where $\theta$ is minimally coupled to gravity,
\begin{equation}\label{eq:alpha1}
S=\int d^{4}x\sqrt{-\bar{g}}\left[\frac{1}{2}\bar{R}+\frac{1}{2}\partial_{\mu}\theta \partial^{\mu}\theta -F\left(\tanh \frac{\theta}{\sqrt{6}}\right)\right].
\end{equation}
Afterwards, one can choose the proper $F$ to get viable inflation models.

Equivalently, we can choose the fixing condition~\cite{Kallosh:2013hoa,Kallosh:2013yoa},
\begin{equation}
\phi^2-\chi^2=1,
\end{equation}
which provides a simple, hyperbolic parameterization for the two scalar fields as 
\begin{equation}
\phi = \cosh \frac{\theta}{\sqrt{6}}, \;\chi = \sinh \frac{\theta}{\sqrt{6}}.
\end{equation}
Then from Eq.~\ref{eq:alpha}, it is straightforward to get
\begin{equation}\label{eq:alpha2}
S=\int d^{4}x\sqrt{-g}\left[\frac{1}{2}R+\frac{1}{2}\partial_{\mu}\theta \partial^{\mu}\theta -F\left(\tanh \frac{\theta}{\sqrt{6}}\right)\right].
\end{equation}
The above action is the same as Eq.~\ref{eq:alpha1}. With the choice of $F(x)\propto x^{2n}$, the authors in Refs.~\cite{Kallosh:2013hoa,Kallosh:2013yoa} have shown that the cosmological predictions are essentially independent on $n$, an attractor behavior in such models. 

However, at this moment it is unclear what a role the approximate SO$(1,1)$ global symmetry plays here. Whether SO$(1,1)$ is essential for the mechanism to have viable inflationary scenarios is not transparent in the above formalism. Below, we shall present a systematic discussion on the general action with local Weyl/conformal symmetry and show that the role of SO$(1,1)$ symmetry is not decisive. 

\subsection{Weyl Gauge symmetry}
We now present the general action with two real scalars $\phi$ and $\chi$ being dynamical fields. Their kinetic terms are in general coupled with Weyl gauge field to maintain the local conformal invariance. The action can be written as follows
\begin{align}\label{eq:weylaction}
S=\int d^{4}x\sqrt{-g}&\left[ \frac{\alpha}{2}\left(\phi^{2}R-6\partial_{\mu}\phi\partial^{\mu}\phi\right)+\frac{\beta}{2}\left(\chi^{2}R-6\partial_{\mu}\chi\partial^{\mu}\chi\right)-V\left(\phi,\chi \right) \right.\nonumber\\
&\left.\;+\frac{\zeta_{1}}{2}D_{\mu}\phi D^{\mu}\phi+\frac{\zeta_{2}}{2}D_{\mu}\chi D^{\mu}\chi-\frac{1}{4g_{W}^{2}}F_{\mu\nu}F^{\mu\nu}\right],
\end{align}
where $\alpha, \beta, \zeta_i$ are numeric parameters, the field strength  $F_{\mu\nu}=\partial_{\mu}W_\nu-\partial_{\nu}W_\mu$, $W_\mu\equiv g_W \omega_\mu$, $\omega_\mu$ is the Weyl gauge field and $g_W$ is the corresponding gauge coupling. The covariant derivative is defined as
\[
D_{\mu}=\left(\partial_{\mu}-W_{\mu}\right).
\]
We emphasize that there is no factor of $i$ in front of Weyl gauge field in the covariant derivative, unlike the usual gauge theory of $U(1)$. As long as the potential has a form as
\begin{equation}
V\left(\phi,\chi \right) = F\left(\frac{\chi}{\phi}\right)\times\left(\sum_{i=0}^{4} \lambda_i \phi^i \chi^{4-i}\right).
\end{equation}
The above action, Eq.~\ref{eq:weylaction}, at classical level is invariant under local conformal transformation Eq.~\ref{eq:weyl1}, together with
\begin{align} W_\mu \rightarrow\overline{W}_{\mu}\left(x\right)=W_{\mu}\left(x\right)-\partial_{\mu}\ln|\lambda\left(x\right)|.
\end{align}

If both $\alpha$ and $\beta$ vanish, Einstein-Hilbert action $R$ would disappear, which goes to a trivial case that is out of our interest. Instead, without losing generality, we shall restrict our discussions with $\alpha > 0$. Then we can always rescale $\phi$ and $\chi$, relabel the parameters or redefine $\zeta_i$ to make $\alpha=1$. In this formalism, $\beta$ can take only three values, $\beta=-1,0,1$, for analytic solutions. It can also be seen immediately that the $\alpha$-attractor is a special case in our formalism with $\beta=-1, \zeta_i=0$ and $\lambda_0=\lambda_4=1,\lambda_2=-2,\lambda_1=\lambda_3=0$.

We shall mainly work with the condition $\phi^2=1$ and later we shall show explicitly the other conditions that can give the same physical models. The action with $\phi^2=1$ can be written as
\begin{align}\label{eq:startl}
S=\int d^{4}x\sqrt{-g}&\left[\left(1+\beta\chi^{2}\right)\frac{1}{2}R +\frac{1}{2}\left(\zeta_{2}-6\beta\right)\partial_{\mu}\chi\partial^{\mu}\chi -V\right.\nonumber\\
&\left.\;+\frac{1}{2}\left(\zeta_{1}+\zeta_{2}\chi^{2}\right)W_{\mu}W^{\mu} -\zeta_{2}W^{\mu}\chi\partial_{\mu}\chi-\frac{1}{4g_{W}^{2}}F_{\mu\nu}F^{\mu\nu}\right]. 
\end{align}

\section{Analytic Solutions}\label{sec:analytic}
The Eq.~\ref{eq:startl} is the Jordan frame action we consider in the following. Now we make a conformal transformation of the metric field
\begin{equation}
\bar{g}_{\mu\nu}\left(x\right)=\Omega^{2}\left(x\right)g_{\mu\nu}\left(x\right),\; \Omega^{2}\equiv1+\beta\chi^{2}, \partial_{\mu}\ln\Omega=\Omega^{-2}\beta\chi\partial_{\mu}\chi.
\end{equation}
The resulting equivalent action with the new metric field can be organized as
\begin{align*}
S =\int d^{4}x\sqrt{-\bar{g}}&\left\{ \frac{1}{2}\left(\bar{R}+\frac{6\beta^2\chi^2}{\Omega^4}\partial_{\mu}\chi \partial^\mu \chi \right)+\frac{\left(\zeta_{2}-6\beta\right)}{2\Omega^{2}}\partial_{\mu}\chi\partial^{\mu}\chi-\text{\ensuremath{\frac{V}{\Omega^{4}}}}\right.\\ &\left.\;+\frac{1}{2\Omega^{2}}\left[\left(\zeta_{1}+\zeta_{2}\chi^{2}\right)W_{\mu}W^{\mu} - \zeta_{2}W^{\mu}\partial_{\mu}\chi^{2}\right]-\frac{1}{4g_{W}^{2}}F_{\mu\nu}F^{\mu\nu}\right\} .
\end{align*} 
Note that the gauge kinetic term $F_{\mu\nu}F^{\mu\nu}$ does not change due to its conformal nature. The gauge-scalar interactions, namely the terms in the bracket of the second line, can be organized as
\begin{align}\label{eq:gaugescalar}
 \frac{\left(\zeta_{1}+\zeta_{2}\chi^{2}\right)}{2\Omega^{2}}\left[W_{\mu}W^{\mu}-W^{\mu}\partial_{\mu}\ln\left(\zeta_{1}+\zeta_{2}\chi^{2}\right)\right]= \frac{\left(\zeta_{1}+\zeta_{2}\chi^{2}\right)}{2\Omega^{2}}\left[w_{\mu}w^{\mu}-\frac{\zeta_{2}^{2}\chi^{2}\partial_{\mu}\chi\partial^{\mu}\chi}{\left(\zeta_{1}+\zeta_{2}\chi^{2}\right)^{2}}\right].
\end{align}
where we have defined the new Weyl gauge field $w_\mu$,
\begin{equation}
	w_\mu = W_\mu - \frac{1}{2}\partial_\mu \ln\left(\zeta_{1}+\zeta_{2}\chi^{2}\right) =W_\mu - \frac{\zeta_{2}\chi\partial_{\mu}\chi}{\zeta_{1}+\zeta_{2}\chi^{2}}.
\end{equation}
This redefinition or gauge transformation does not change the kinetic term for $w_\mu$, $F_{\mu\nu}F^{\mu\nu}$, but contributes to the kinetic term for $\chi$, as shown in Eq.~\ref{eq:gaugescalar}.

Now we can present the total kinetic term of $\chi$ for general $\zeta_{i}$, $\dfrac{1}{2}K(\chi) \partial_{\mu}\chi \partial^\mu \chi$. The coefficient $K(\chi)$ is collected as the sum of three contributions,
\begin{align}
K(\chi) =  \frac{6\beta^2\chi^2}{\Omega^4} +\frac{\left(\zeta_{2}-6\beta\right)}{\Omega^{2}} -\frac{\zeta_{2}^{2}\chi^{2}}{\Omega^{2}\left(\zeta_{1}+\zeta_{2}\chi^{2}\right)}
=  \frac{\zeta_{1}\zeta_{2}\left(1+\beta\chi^{2}\right) - 6\beta\left(\zeta_{1}+\zeta_{2}\chi^{2}\right)}
{\Omega^{4}\left(\zeta_{1}+\zeta_{2}\chi^{2}\right)}.
\end{align}
As long as $K(\chi) >0$, we can make the kinetic term canonical by defining a new field variable $\theta(\chi)$ through
\begin{equation}\label{eq:k1}
\frac{d\theta }{d\chi}=\sqrt{K(\chi)}. 
\end{equation}
Once obtaining the canonical kinetic term, the full action is the following
\begin{equation}\label{eq:can}
S=\int d^{4}x\sqrt{-\bar{g}}\left[\frac{1}{2}\bar{R}+\frac{1}{2}\partial_{\mu}\theta\partial^{\mu}\theta-\frac{V\left(\theta \right)}{\Omega^{4}}-\frac{1}{4g_{W}^{2}}F_{\mu\nu}F^{\mu\nu}+\frac{\zeta_{1}+\zeta_{2}\chi^{2}(\theta)}{2\Omega^{2}}w_\mu w^\mu \right],
\end{equation}
which describes the Einstein's gravity coupled with a scalar field $\theta$ and a massive vector $w_\mu$. $\theta$ has a potential $V/\Omega^{4}$ and couples to $w_\mu$ once we expand the last term in the bracket.

For general $\zeta_i$, there is no compact analytical solution for the above differential equation, Eq.~\ref{eq:k1}. However, in some special cases, we have found very simple analytic relations, tabulated in Table.~\ref{tab:cases}. For example, if $\beta\neq 0$ and $\zeta_2=\beta \zeta_1$, we have 
\begin{equation}\label{eq:k2}
\sqrt{K(\chi)}=\frac{\sqrt{\beta\left(\zeta_{1}-6\right)}}{\left(1+\beta\chi^{2}\right)}.
\end{equation}  
Then, for $\beta=-1$, we can obtain
\begin{align*}
\frac{d\theta}{d\chi} & =\frac{\sqrt{6-\zeta_1}}{1-\chi^{2}}\Rightarrow \chi = \tanh \frac{\theta}{\sqrt{6-\zeta_1}}.
\end{align*}
When $\zeta_1=0$, this result fully agree with the $\alpha-$attractor case. From this calculation, we can also get the theoretically allowed domain $\zeta_1<6$ from the right sign of the kinetic term.

For $\beta=1$, we have 
\[
\frac{d\theta}{d\chi}=\frac{\sqrt{\zeta_1-6}}{1+\chi^{2}} \Rightarrow \chi = \tan \frac{\theta}{\sqrt{\zeta_1-6}}.
\]
In such a case, as long as $\zeta_1>6$, we can have a consistent theory with a normal scalar field $\theta$ and the starting action can have an approximate SO$(2)$ global symmetry. This illustration also explains why in the case of $\zeta_i=0$, $\beta$ has to be $-1$ (the resulting SO$(1,1)$ symmetry is reached). Otherwise we would get a wrong sign for the kinetic term of $\theta$.

In the above two cases, we have 
\begin{equation}\label{eq:mass}
\frac{\zeta_{1}+\zeta_{2}\chi^{2}(\theta)}{2\Omega^{2}}w_\mu w^\mu=\frac{\zeta_1}{2}w_\mu w^\mu,
\end{equation}
which indicates that $\theta$ actually decouples from Weyl gauge field $w_\mu$ whose mass is given by $m_w=g_W\sqrt{\zeta_1}M_P$. 

\begin{table}
	\centering
	\footnotesize
	\begin{tabular}{|c|c|c|c|}
		\hline 
		$\alpha=1$ & $K\left(\chi\right)$ & $\theta=\theta\left(\chi\right)$ & $\chi = \chi \left(\theta \right)$
		\tabularnewline
		\hline 
		{$ \begin{aligned}
			&\beta=0, \\
			\zeta_{1}\zeta_{2}\neq 0,&\dfrac{\zeta_{2}}{\zeta_{1}}=C
			\end{aligned}$} & $\dfrac{C\zeta_{1}}{1+C\chi^{2}}$ & 
		{$ \begin{aligned}
			C>0,\zeta_{1}>0 & \Rightarrow \theta=\sqrt{\zeta_{1}}\textrm{arcsinh}\left(\sqrt{C}\chi\right) \\ 
			C<0,\zeta_{1}<0 & \Rightarrow \theta=\sqrt{-\zeta_{1}}\arcsin\left(\sqrt{-C}\chi\right)\\
			C<0,\zeta_{1}>0 & \Rightarrow \theta=\sqrt{\zeta_{1}}\textrm{arccosh}\left(\sqrt{-C}\chi\right) 
			\end{aligned} $}
		&
		{$ 	\begin{aligned}
			\chi  & =\frac{1}{\sqrt{C}}\sinh\dfrac{\theta}{\sqrt{\zeta_{1}}}\\
			\chi  & =\frac{1}{\sqrt{-C}}\sin\dfrac{\theta}{\sqrt{-\zeta_{1}}}\\
			\chi  & =\frac{1}{\sqrt{-C}}\cosh\dfrac{\theta}{\sqrt{\zeta_{1}}}
			\end{aligned}	$}
		\tabularnewline
		\hline 
		{$	\begin{aligned}
			\zeta_{1} & =0, \\
			\textrm{ or/and }\zeta_{2}&=0
			\end{aligned} 
			$}
		& $\dfrac{-6\beta}{\left(1+\beta\chi^{2}\right)^{2}}$ & $\beta=-1\Rightarrow\theta=\dfrac{\sqrt{6}}{2}
		\ln\dfrac{1+\chi}{1-\chi} $ & {$ \begin{aligned}
			&\chi=\tanh\dfrac{\theta}{\sqrt{6}}
			\end{aligned}$}
		\tabularnewline
		\hline 
		{$ \begin{aligned}
			&\beta=+1,\\ 
			\zeta_{1}=&+\zeta_{2}\equiv\zeta
			\end{aligned} $} & $\dfrac{\zeta-6}{\left(1+\chi^{2}\right)^{2}}$ & $\zeta>6\Rightarrow\theta=\sqrt{\zeta-6}\arctan\chi$ & $\chi=\tan\dfrac{\theta}{\sqrt{\zeta-6}}$
		\tabularnewline
		\hline 
		{$ \begin{aligned}
			&\beta=-1,\\ 
			\zeta_{1}=&-\zeta_{2}\equiv\zeta
			\end{aligned} $}  & $\dfrac{6-\zeta}{\left(1-\chi^{2}\right)^{2}}$ & $\zeta<6\Rightarrow\theta=\dfrac{\sqrt{6-\zeta}}{2}
		\ln\dfrac{1+\chi}{1-\chi} $ & {$ \begin{aligned}
			&\chi=\tanh\dfrac{\theta}{\sqrt{6-\zeta }}
			\end{aligned}$}
		\tabularnewline
		\hline 
	\end{tabular}
	\caption{The analytic solutions of $\theta\left(\chi\right)$ or $\chi \left(\theta \right)$ for different $\beta, \zeta_1 $ and $\zeta_2$. As far as we have investigated, these are the only cases that allow a transparent and compact solution.  \label{tab:cases}}
\end{table}

There is no apparent global symmetry in the cases of the first two rows in Table.~\ref{tab:cases}, except the one with $\zeta_i=0, \beta =-1$, which is exactly the original $\alpha$-attractor. However, if one can allow the additional explicit breaking of some global symmetry from the Ricci scalar term, then the cases of the first row exhibit approximate SO(1,1) and SO(2) symmetry when $C=-1$ and $C=1$, respectively. Note that this additional breaking should not be interpreted as a minor modification to the cases with SO(1,1) and SO(2) symmetry. In fact, for the same potential $V$, the modification of inflation could be significant. Let us take $\beta=0, \zeta_1=1, C=-1$ as an example in comparison with $\beta=-1$ case. For concrete discussion, we may choose $V=(\chi/\phi)^n(\chi^2-\phi^2)^2$ as in $\alpha$-attractor. Then we would have $V(\theta)=\tanh^n\theta/\sqrt{5}$ for $\beta=-1$ but $V(\theta)=\cosh^n\theta \sinh^4\theta$ for $\beta=0$. These two potentials have very different behaviors which result in different inflation dynamics. The potential for $\beta=-1$ is flat at large $\theta$ since $\tanh\theta/\sqrt{5}\rightarrow 1$ as $\theta\rightarrow+\infty$, hence viable inflation can happen for positive $n$. While the potential for $\beta=0$ goes as $e^{(4+n)\theta}$ and slow-roll conditions can not be satisfied at large $\theta$ for positive $n$.

We emphasize that the above discussions by no means indicate that viable inflation is not possible in the cases with $\beta=0$. It just means that $\beta=0$ can not be treated as a small correction to the cases with $\beta=\pm 1$. The main reason behind is the different asymptotic behavior of the solution $\chi(\theta)$ in different cases, which require the proper choice of the potential $V$ for successful inflation. For instance, if we choose $V=(\chi/\phi)^{-4}(\chi^2-\phi^2)^2$ for the $\beta=0$ case, we would get $V(\theta)=\tanh^4\theta$. This potential is very flat at large $\theta$ and can give viable inflation.

Now we explicitly demonstrate under what circumstances, the condition, $\phi^{2}+\beta\chi^{2}=1$, can give the equivalent final theory. For $\beta=0$, it simply reduced the above case. For $\beta=\pm 1$, we can use the parametrization 
\begin{align}
\phi=&\cos\theta/\sqrt{\zeta_1-6},\;\chi=\sin\theta/\sqrt{\zeta_1-6},\;\textrm{for }\beta=\alpha=1, \\
\phi=&\cosh\theta/\sqrt{6-\zeta_1},\;\chi=\sinh\theta/\sqrt{6-\zeta_1},\;\textrm{for }\beta=-\alpha=-1.
\end{align}
Since there is no transparent form for the general $\zeta_i$ case, we illustrate with $\zeta_1=\zeta_2$ for $\beta=\alpha=1$, and $\zeta_1=-\zeta_2$ for $\beta=-\alpha=-1$.  
Using the above parametrization, we can perform the calculations straightforwardly and obtain in both cases 
\begin{equation}\label{eq:can2}
S=\int d^{4}x\sqrt{-g}\left[\frac{1}{2}R + \frac{1}{2}\partial_{\mu}\theta\partial^{\mu}\theta -V\left(\theta \right)-\frac{1}{4g_{W}^{2}}F_{\mu\nu}F^{\mu\nu}+\frac{\zeta_{1}}{2}w_\mu w^\mu\right].
\end{equation}
There is a crucial difference in the final potential where the factor $\Omega^{-4}$ appears in Eq.~\ref{eq:can} but not in 
Eq.~\ref{eq:can2}. This leads us to the observation that the above formula agrees with Eq.~\ref{eq:can} only if the potential can be factorized into the form where one factor also respects the global symmetries as the kinetic terms, SO$(2)$ or SO$(1,1)$, which means
\begin{equation}
 V(\phi,\chi) = F\left(\frac{\chi}{\phi}\right)\times\begin{cases}
 \left(\phi^2+\chi^2\right)^2, & \beta=+1,\zeta_1=+\zeta_{2}, \\
 \left(\phi^2-\chi^2\right)^2, & \beta=-1,\zeta_1=-\zeta_{2}.
 \end{cases}
\end{equation}
Having this form, the factor $\left(\phi^2+\beta\chi^2\right)^2$ in the potential cancels with $\Omega^{-4}$ from the conformal transformation, and the potential for $\theta$ is 
\begin{equation}
 F\left(\frac{\chi}{\phi}\right)=\begin{cases}
F\left(\tan \dfrac{\theta}{\sqrt{\zeta_1-6}}\right), & \beta=+1,\zeta_1=+\zeta_{2}, \\
F\left(\tanh \dfrac{\theta}{\sqrt{6-\zeta_1}}\right), & \beta=-1,\zeta_1=-\zeta_{2}.
\end{cases}
\end{equation}

In general, when $\zeta_2 \neq \pm \zeta_1$, our calculations exhibit that the symmetry breaking conditions, $\phi^2=1$ and $\phi^{2}+\beta\chi^{2}=1$, would give different potentials for the field $\theta$. This result has some similarity with the Higgs mechanism in particle physics, where the physical theories also depend on how the gauge symmetries are broken by the different vacuum configurations of the Higgs fields.

The above discussions can be generalized to multi-field cases, $\phi_i(i=1,2,...,k)$. We can normalize the fields with the corresponding $\beta_i=\pm 1$ ($l$ positive $\beta_i$ and $m$ negative ones with $l+m=k-1$). The results would imply that SO$(l+1,m)$ is the approximate global symmetry. The parameterization of fields would be straightforward and involve high-dimensional spherical coordinates.

\section{Phenomenology}\label{sec:pheno}

\subsection{Inflation}

The analytic solutions we obtained in Table.~\ref{tab:cases} can be classified into two categories, trigonometric functions and hyperbolic ones. Since we may choose $F(\chi/\phi)$ at will, any solution in each category can be representative. The hyperbolic solutions have been extensively discussed in the literature as $\alpha$-attractor~\cite{Kallosh:2013hoa, Kallosh:2013yoa}, so we do not repeat the analysis here. Instead, we discuss the trigonometric ones and illustrate with one class of inflationary models. 

For concreteness, we choose the solution in the case of $\beta=1$ and $\zeta_1=\zeta_2=\zeta$ with an approximate SO$(2)$ symmetry. Then we have $\chi=\tan \theta/\sqrt{\zeta-6}.$ As long as the functions of $F(\tan \theta/\sqrt{\zeta-6})$ can satisfy the slow-roll conditions, inflation can happen. For a proof-of-principle example, in the following we discuss the inflationary observations with the following function $F$ or potential,
\begin{equation}\label{eq:infmodel}
F(x)=\lambda \left(\frac{1}{1+x^2}\right)^{{1}/{n}} \Rightarrow F[\theta] = \lambda \left[\cos^{2}(A{\theta})\right]^{{1}/{n}}, \; A\equiv 1/{\sqrt{\zeta-6}},
\end{equation}
where $n$ is a parameter of the chosen potentials. Note that $F[\theta]\geq 0$ and the potential minimum is reached when $A\theta=\pi/2$ (we only consider the first period, $A\theta\subset[0,\pi]$). For $n=1$, the potential has the same form as the one in natural inflation~\cite{Freese:1990rb}. This also indicates that polynomial potentials in Jordan frame can induce trigonometric potentials in Einstein frame, providing an alternative origin of cosine-like inflation.

The slow-roll parameters are calculated as
\begin{align}\label{eq:slowroll}
\epsilon &= \frac{1}{2}\left(\frac{F_\theta }{F}\right)^2=\frac{2 A^2 \tan ^2(A \theta)}{n^2},\\
\eta &= \frac{F_{\theta \theta}}{F}=-\frac{2 A^2 \sec ^2(A \theta) \left[\cos (2 A \theta)+n-1\right]}{n^2},
\end{align}
where $F_\theta \equiv dF[\theta]/d\theta$ and $F_{\theta\theta}\equiv dF_\theta/d\theta$. 
The observable scalar index $n_s$ of the power spectrum and tensor-to-scalar ratio $r$ for the signal strength of primordial gravitational wave are determined by $n_s=1-6\epsilon+2\eta$ and $r=16\epsilon$,
\begin{equation}\label{eq:nsr}
n_s=1 - \frac{2 A^2 \sec ^2(A \theta) \left[1+2n-\cos (2 A \theta)\right]}{n^2},\ r = \frac{32 A^2 \tan ^2(A \theta)}{n^2}.
\end{equation}
To solve the flatness problem, the early universe should have expanded with enough e-folding number $N\sim [50,60]$ before inflation ends,
\begin{equation}\label{eq:efold}
N\equiv \ln \frac{a_{e}}{a_{i}}\simeq \int^{t_\textrm{end}}_t Hdt \simeq \int ^{\theta_i}_{\theta_\textrm{e}}\frac{d\theta}{\sqrt{2\epsilon}}, 
\end{equation}
where $a_i (a_e)$ is the scale factor at initial (end) time of the inflation, $\theta_i (\theta_e)$ is the corresponding field value, and $H$ is the Hubble parameter. Here $\theta_e$ is determined by the violation of slow-roll condition, $\epsilon\sim 1$ or $\eta\sim 1$. 

\begin{figure}[t]
	\includegraphics[width=0.65\textwidth,height=0.52\textwidth]{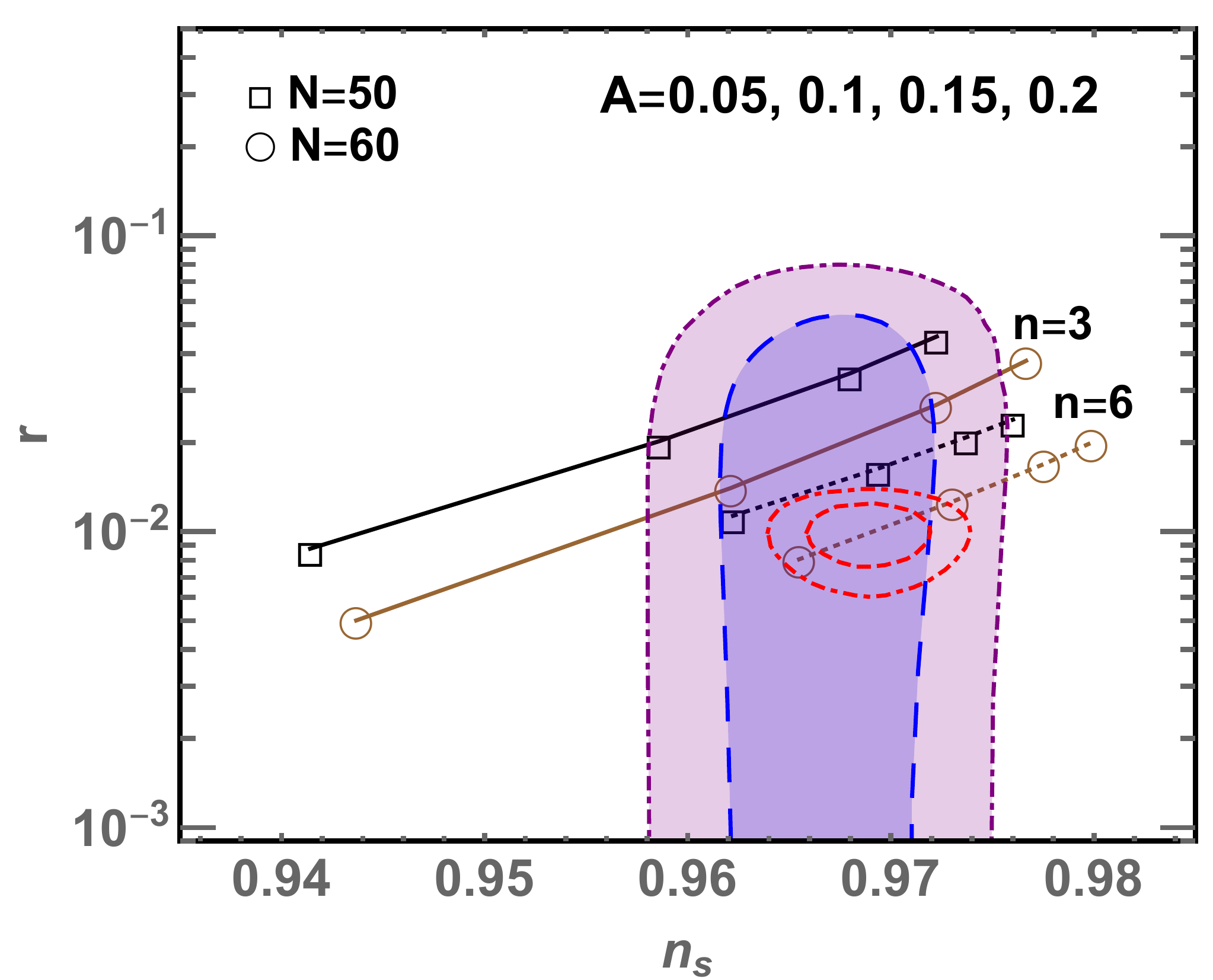}
	\caption{Illustration of $(n_s, r)$ for $A=0.05,0.10,0.15,0.2$ (from right to left) when $n=3,6$. The two solid lines and two dotted lines are for $n=3$ and $n=6$ respectively. The calculated values of $(n_s, r)$ with 4 different $A$s are shown for e-folding number $N=50$ (squares) and $60$ (circles), in comparison with the shaded regions allowed by {\it Planck}~\cite{Planck:2018} with 1-$\sigma$ (blue) and 2-$\sigma$ (purple), and the future projection of CMB-S4~\cite{Abazajian:2016yjj} in red smaller contours.
		\label{fig:r-ns}}
\end{figure}

For small $n$ and $2A^2N/n\ll 1$, we have the following approximate formula for $n_r$ and $r$,
\begin{equation}\label{eq:nsra}
n_s\approx 1-\frac{2}{N}\left(1+\frac{2A^2N}{n}\right), r \approx  \frac{16}{Nn}\left(1-\frac{2A^2N}{n}\right),
\end{equation}
which are useful for qualitative understanding. For instance, we would expect both $n_s$ and $r$ should decrease as $A^2$ increases, which will be reflected in Fig.~\ref{fig:r-ns} where we employ the precise estimation. For precision calculation, we numerically solve the Eq.~\ref{eq:efold} with the boundary conditions, $N(\theta_e)=0$ and $N(\theta_i)=50$ or $60$. Once having the value of $\theta_i$, we put it into Eq.~\ref{eq:nsr} and obtain $n_s$ and $r$. 

In Fig.~\ref{fig:r-ns} we illustrate the cases with $n=3, 6$ and show the theoretical predictions of $(n_s, r)$ for $A=0.05,0.10,0.15,0.2$. The solid line that connects 4 squares (circles) represents the values of $(n_s, r)$ with $N(\theta_i)=50$ (60) when $n=3$, while the dotted lines are for $n=6$. We also contrast our predictions with the latest constraints from {\tt Planck}~\cite{Planck:2018} (color-shaded regions) and the future projection of the next generation CMB experiments~\cite{Abazajian:2016yjj} (two smaller contours). It is seen that the proposed model in Eq.~\ref{eq:infmodel} can be consistent with current observations and will be probed by future CMB experiments.  

For $n=1$ or natural inflation, we have verified that it has already been excluded by {\tt Planck}~\cite{Planck:2018} more than $2\sigma$. For $n=2$, we have also checked almost all the predictions are out of $2\sigma$ region. For $n\geq 3$, our scenario is viable. For larger $n$, $(n_s,r)$ would be shifted downwards to the right and the effects can be partially compensated by increasing $n$, which can be understood from Eq.~\ref{eq:nsra} and seen in Fig.~\ref{fig:r-ns}. In general, larger $n$ would give smaller $r$. 

The parameter $A\sim 0.1$ implies $\zeta\sim 100$ from the relation $A\equiv 1/\sqrt{\zeta - 6}$. At first sight, $\zeta\sim 100$ might seem a large number. However, this is because we normalized $\alpha=\beta=1$ in the Lagrangian, Eq.~\ref{eq:weylaction}. If we keep both $\alpha$ and $\zeta$ general from the start, we shall find $A=\sqrt{\frac{\alpha}{\zeta-6\alpha}}$. Then we would get $\zeta \sim 10$ for $\alpha=0.1$ and $\zeta=1$ for $\alpha\sim 0.01$, which should be acceptably natural. This model belongs to the large-field inflations since the evolved field value $\Delta \theta > M_p$, but the energy scale at inflation is around $10^{16}\GeV$.

We would like to make a brief discussion about the reheating process after inflation. When the slow-roll conditions are violated, the exponential expansion stops and the inflation field oscillates around the potential minimum, $\theta_0 = \pi/(2A)$. And the universe enters the matter-dominated era. For perturbative reheating, one may introduce interactions between $\chi$ and other fermions $\psi$ or scalars $s$, such as $\chi \bar{\psi}\psi$ and $\chi s^3$, which preserve the conformal symmetry but break in general the global SO$(2)$ symmetry. The new interactions would make $\theta$ decay and transfer its energy into radiation. So that our universe is radiation-dominant after reheating and can have a successful nucleosynthesis.

\subsection{Weyl Gauge Boson}
We notice that in all cases there is a $Z_2$ symmetry $w_\mu\rightarrow -w_\mu$ for Weyl gauge boson $w_\mu$. If $w_\mu$ particles were produced in the early universe, there could leave some relic at present.

As shown in Eq.~\ref{eq:mass} or Eq.~\ref{eq:can2}, when there is an approximate global SO(1,1) or SO(2) symmetry, $w_\mu$ is essentially decoupled from the inflation field $\theta$ and only interacts gravitationally. In such cases, $w_\mu$'s mass is given by $m_w=g_W \sqrt{\zeta} M_p$. If $g_W^2\zeta\sim 1$, $w_\mu$'s mass is around Planck scale and $w_\mu$ would be too heavy to be produced in the early universe after inflation. Hence, there is no observational problem. On the other hand, if $g_W^2\zeta$ is small enough, $w_\mu$ can be produced gravitationally and actually be a dark matter candidate~\cite{Tang:2019uex}, whose relic density is given by~\cite{Graham:2015rva, Ema:2019yrd},
\begin{equation}
\Omega_w \simeq 0.25\times \sqrt{\dfrac{m_w}{6\times 10^{-11}\GeV}}\times \left(\dfrac{H}{10^{13}\GeV}\right)^2.
\end{equation}	
If we require $m_w\leq H$ in an inflation model, we would get an upper bound $m_w\lesssim 10^9\GeV$ when the equality is reached. One may expect that $w_\mu$ as dark matter is subjected to isocurvature perturbation constraint since its longitudinal mode is similar to a scalar degree of freedom. However, there is a crucial difference for vector dark matter produced from inflationary fluctuation~\cite{Graham:2015rva, Ema:2019yrd}. Although they can be copiously produced in the early universe, the main contribution lies around the modes with characteristic wave number $k\sim a m_w$ ($a$ is the scale factor at horizon crossing), which is concentrated around small scales. The perturbations for the large or CMB scales still inherit from inflaton and is adiabatic. The fundamental reason is that the equation of motion for the longitudinal mode of a massive vector,
\begin{equation}\label{eq:eom}
\left(\partial_{t}^{2}+\frac{3 k^{2}+a^{2} m_w^{2}}{k^{2}+a^{2} m_w^{2}} H \partial_{t}+\frac{k^{2}}{a^{2}}+m_w^{2}\right) w_{L}(k)=0,
\end{equation}
has different behavior from that for a scalar $\varphi$, $\left(\partial_{t}^{2}+3 H \partial_{t}+k^{2}/a^{2}+m_\varphi^{2}\right) \varphi(k)=0$, see Refs.~\cite{Graham:2015rva, Ema:2019yrd} for detailed investigations. 

In the other cases without global symmetry, there can be a direct coupling between inflaton $\theta$ and Weyl gauge boson, originating from the term $\dfrac{\zeta_{1}+\zeta_{2}\chi^{2}(\theta)}{2\Omega^{2}}w_\mu w^\mu$. During inflation, we would get an effective mass for Weyl gauge boson, $m^2_{\textrm{eff}}=[\zeta_{1}+\zeta_{2}\chi^{2}(\theta)]/\Omega^{2}$. One immediate conservative constraint we can get is from the requirement $m^2_{\textrm{eff}}>0$ to avoid tachyon, namely,
\begin{equation}
\frac{\zeta_{1}+\zeta_{2}\chi^{2}(\theta)}{1+\beta \chi^{2}(\theta)}>0.
\end{equation}
Another effect is that time-changing mass usually induces particle production, then $w_\mu$'s relic abundance in turn would put a constraint on $\chi^{2}(\theta)$. Since this production depends on the  evolution of the inflaton field or inflation models, a full discussion would be beyond our scope here.

\section{Conclusion}\label{sec:conclusion}
We have presented a systematic analysis on the inflation models with local conformal symmetry, together with the Weyl gauge field. One of our motivations is to understand why SO$(1,1)$ plays a so special role in the appealing $\alpha$-attractor model. We have found that the underlying reason is the positivity of the kinetic term for the inflaton field. Moreover, within the general formalism in the presence of Weyl gauge field, we have identified the other viable symmetry, SO$(2)$. We have also tabulated in table.~\ref{tab:cases} all the possible analytic solutions that relate the inflaton fields between Jordan frame and Einstein frame. These solutions can be classified into two categories, trigonometric functions and hyperbolic ones. Finally, we have demonstrated a class of inflation models with an approximate SO$(2)$ global symmetry and shown it can be consistent with the latest cosmological observations and will be probed by future CMB experiments.

$\\$
\begin{Large}
\textbf {Acknowledgments}
\end{Large}$\\$
YT is supported by National Science Foundation of China (NSFC) under Grants No.~11851302. YLW is supported in part by NSFC under Grants No.~11851302, No.~11851303, No.~11690022, No.~11747601, and the Strategic Priority Research Program of the Chinese Academy of Sciences under Grant No. XDB23030100 as well as the CAS Center for Excellence in Particle Physics (CCEPP).

%\bibliographystyle{../utphysMa}
%\bibliographystyle{../notitle}
%\bibliographystyle{../JHEP}
%\bibliographystyle{apsrev4-1}
%\bibliography{references}

%

\end{document}